\documentclass[12pt,preprint]{aastex}
\usepackage{natbib}

\newcommand{\etal}{et al.\thinspace}

\slugcomment{}
\shorttitle{}
\shortauthors{Sudilovsky, Smith, Savaglio}

\begin{document}

\title{Dusty MgII Absorbers: Implications for the GRB/Quasar Incidence Discrepancy}

\author{Vladimir Sudilovsky\altaffilmark{1,2}}
\author{Donald Smith\altaffilmark{1}}
\author{Sandra Savaglio\altaffilmark{2}}

\altaffiltext{1}{Physics Department, Guilford College, 5800 Friendly Ave, Greensboro, North Carolina 27410} 
\altaffiltext{2}{Max-Planck-Institut f\"ur Extraterrestrische Physik, Giessenbachstrasse, D-85748 Garching bei M\"unchen, Germany}

\begin{abstract}
There is nearly a factor of four difference in the number density of intervening MgII absorbers as determined from gamma-ray burst (GRB) and quasar lines of sight. We use a Monte-Carlo simulation to test if a dust extinction bias can account for this discrepancy. We apply an empirically determined relationship between dust column density and MgII rest equivalent width to simulated quasar sight-lines and model the underlying number of quasars that must be present to explain the published magnitude distribution of SDSS quasars. We find that an input MgII number density $dn/dz$ of $0.273\pm0.002$ over the range $0.4 \le z \le 2.0$ and with MgII equivalent width $W_0 \ge 1.0 \mathrm{\AA}$ accurately reproduces observed distributions. From this value, we conclude that a dust obstruction bias cannot be the sole cause of the observed discrepancy between GRB and quasar sight-lines: this bias is likely to reduce the discrepancy only by $\sim 10\%$.
\end{abstract}


\section{Introduction} \label{introduction} 
The Sloan Digital Sky Survey (SDSS) has greatly expanded our knowledge of the large scale structure of the universe, in large part by greatly increasing the number of known quasars at high redshift to $\sim 10^5$ \citep{PETERSON, SDSS_DR6_PAPER}. These bright, distant sources of light also provide a means to probe intergalactic space by allowing us to identify and analyze spectral absorption lines caused by intervening matter. These measurements can constrain the star formation rate history \citep{HAMANN1999, HOPKINS2006}, the formation of large scale structure \citep{CLOWES1999, CABANAC2005}, the evolution of chemical enrichment in the high-redshift universe \citep{BAKER2000, PETTINI2001, MEHLERT2003, PROCHASKA2007}, and the time of re-ionization \citep{FAN2006, MESINGER2007}. 

These investigations are only made possible by a population of bright sources at high redshifts. Quasars provide such a population, but so do the brief flares of Gamma-Ray Burst (GRB) sources. The catalog of quasar observations is much larger than the sample of GRBs that have been studied with optical spectroscopy, in large part due to the transient nature of GRBs. GRBs are first detected in gamma-rays, thus they are therefore not subject to the same observational biases as quasars. For this reason, the GRB population plays a potentially crucial role in the study of intervening absorption systems.

\citet{PROCHTER} used GRBs to study absorption systems that lie along the lines of sight. They found on average four times the number of intervening strong (equivalent width greater than one angstrom) MgII absorbers in GRB sight-lines per unit redshift as compared to the number of MgII absorbers in quasar sight-lines. This result is surprising because both populations sample random, independent sight-lines. Possible factors that may explain this discrepancy include differing MgII covering factors in GRB and quasar beams \citep{PONTZEN2007, FRANK2007}, weak gravitational lensing affecting a population of absorbers, a dust extinction bias, and host-associated MgII absorbers being incorrectly identified as intervening systems \citep{PROCHTER, PORCIANI2007}. \citet{SUDILOVSKY} surveyed intervening CIV systems and found no difference in their number density in either type of sight-line, which is also confirmed by \citet{TEJOS2007}. \citet{SUDILOVSKY} argued that the simplest explanation for this discrepancy lies in a selection bias. Samples of quasars that are detected based on optical magnitudes may be biased towards sight-lines that contain fewer intervening MgII systems, because MgII systems are tracers of dense, potentially dusty clouds, which would attenuate optical light passing through them \citep{MENARD2005,MENARD2008,YORK2006,KHARE2005dust}. 

We report on the results of a Monte-Carlo simulation to test the selection bias introduced by dust extinction on optically limited surveys. We simulated a population of quasars that follow distributions in redshift and magnitude determined empirically from the SDSS. We applied a magnitude shift to each quasar that corresponds to the effect of the dust associated with simulated MgII absorbers placed along that quasar's line of sight. For various input MgII number densities $dn/dz$\footnote{For the remainder of this paper, we constrain the quantity $dn/dz$ to $0.4 \le z \le 2.0$ and MgII equivalent width $W_0 \ge 1.0 \mathrm{\AA}$}, we added quasars to the simulation until the number of quasars brighter than the SDSS's limiting magnitude matched the number of quasars in the actual SDSS catalog. Finally, we determine the input $dn/dz$ for which the simulation returns the SDSS-observed distributions of MgII $dn/dz$ and quasar $i$-magnitude. Using this technique, we determine if dust can explain the GRB and quasar MgII number density discrepancy.

In this paper, we explain the observationally-derived probability distributions upon which our simulation is based in \S \ref{data}. In \S \ref{methods} we present the algorithm used by our simulation, and in \S \ref{results} we present the statistical properties extracted from many simulation runs. \S \ref{discussion} places these results in the context of current research.

\section{Observational Data} \label{data} 

\subsection{The MgII Sample} \label{data:mgii_sample} 

MgII absorption is readily detected in the optical spectra of objects with $0.4 \lesssim z \lesssim 2.2$ \citep*{LANZETTA1987}. MgII is found at small galaxy impact parameters and can therefore be used as a tracer of galaxies \citep{LANZETTA1990,STEIDEL1994}. The optical depth of dust has been well approximated over $0.4 \le z \le 2.0$ and MgII equivalent width $W_0 \ge 1.0 \mathrm{\AA}$ with a power law of the form $\langle \tau_{v} \rangle(W_0) = \tau_{v,0} \left(\frac{W_0}{1\rm{\AA}}\right)^{\alpha}$, where $\alpha = 1.88\pm0.17$,  $\tau_{v,0}=(2.5\pm0.2)\times 10^{-2}$, and $W_0$ is the rest equivalent width of the $\lambda2796$ feature
\citep{MENARD2008}.

We calculated the extinction caused by dust associated with MgII absorbers as a function of their equivalent width $W_0$, their redshift, and their abundance along sight-lines to distant objects such as quasars or GRBs. \citet{MENARD2008} determined the correlation between MgII $W_0$ and the optical depth of dust, and the dust optical depth yields the extinction calculation for one absorber system. We limited our simulated absorbers to $W_0 \ge 1$ \AA\ to remain consistent with the domain of applicability of the MgII number density and $W_0$ distribution\footnote{Also, evidence suggests that dust extinction is insignificant below this cutoff \citep{YORK2006,MENARD2008}}. We use the $W_0$ distribution empirically derived by \citet*{NESTOR2005}. Since it is possible that there exist a population of strong MgII absorbers that contain so much dust that they simply cannot be observed through optical spectroscopy, our estimate of the fraction of dust-obscured quasars should be regarded as a lower limit. We use the MgII $dn/dz$ as determined by \citet{PROCHTER} as an upper limit in the simulation.

\subsection{The Quasar Sample} 

We used the main spectroscopic quasar sample from the SDSS fifth data release (DR5) as our control catalog. The DR5 differentiates quasars from stars by color and redshift criteria according to an algorithm described by \citet{SDSS_TECHPAPER}. The SDSS main quasar sample consists of quasars in the magnitude range $15.0 < i < 19.1$.  We exclude quasars with $i > 19.1$ from our analysis. These quasars are detected by the SDSS serendipity fibers and are therefore not detected using a well-defined search algorithm. 

\begin{figure}
    	\begin{centering}
     	\includegraphics[scale = .5]{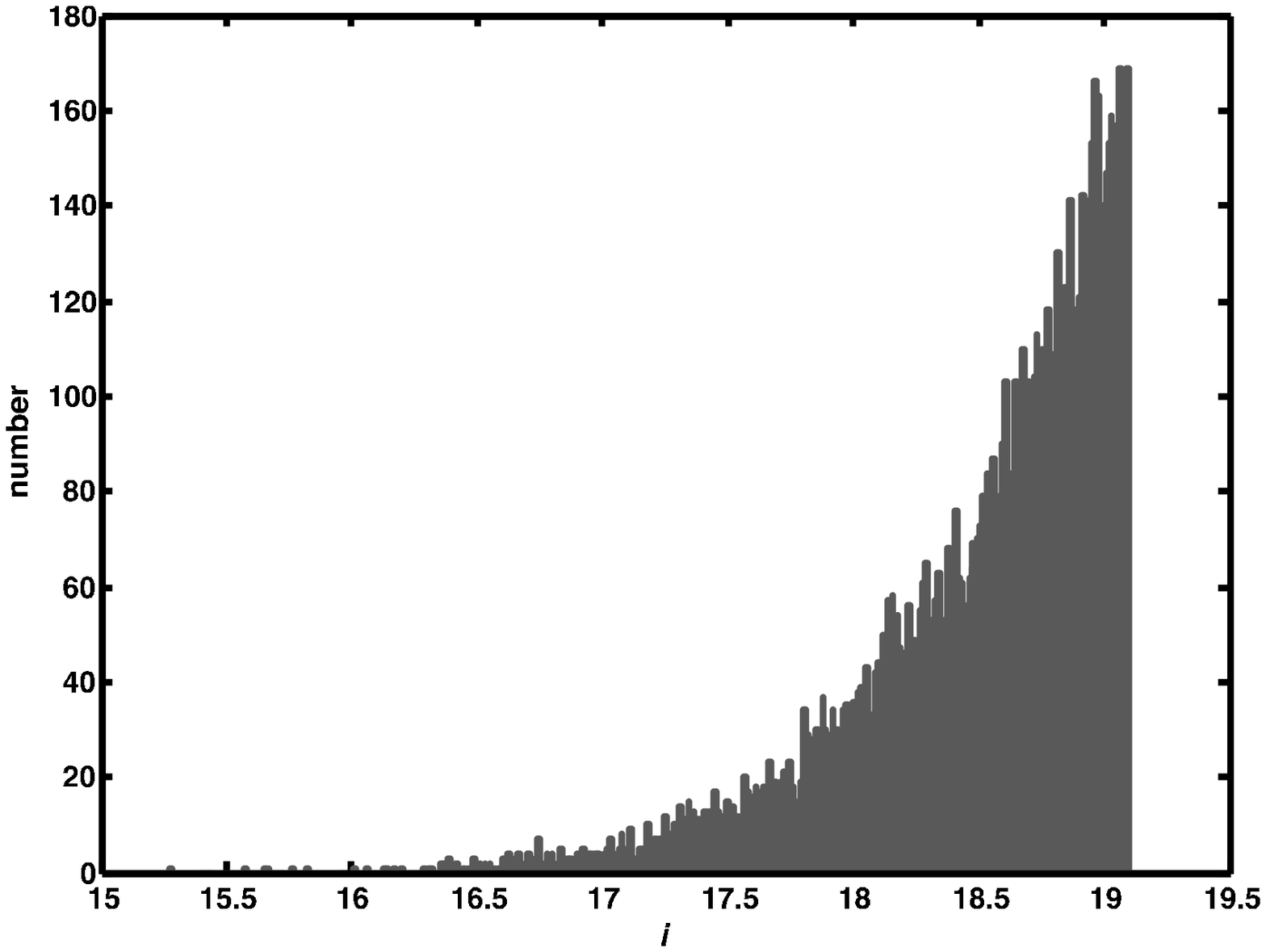}\includegraphics[scale = .5]{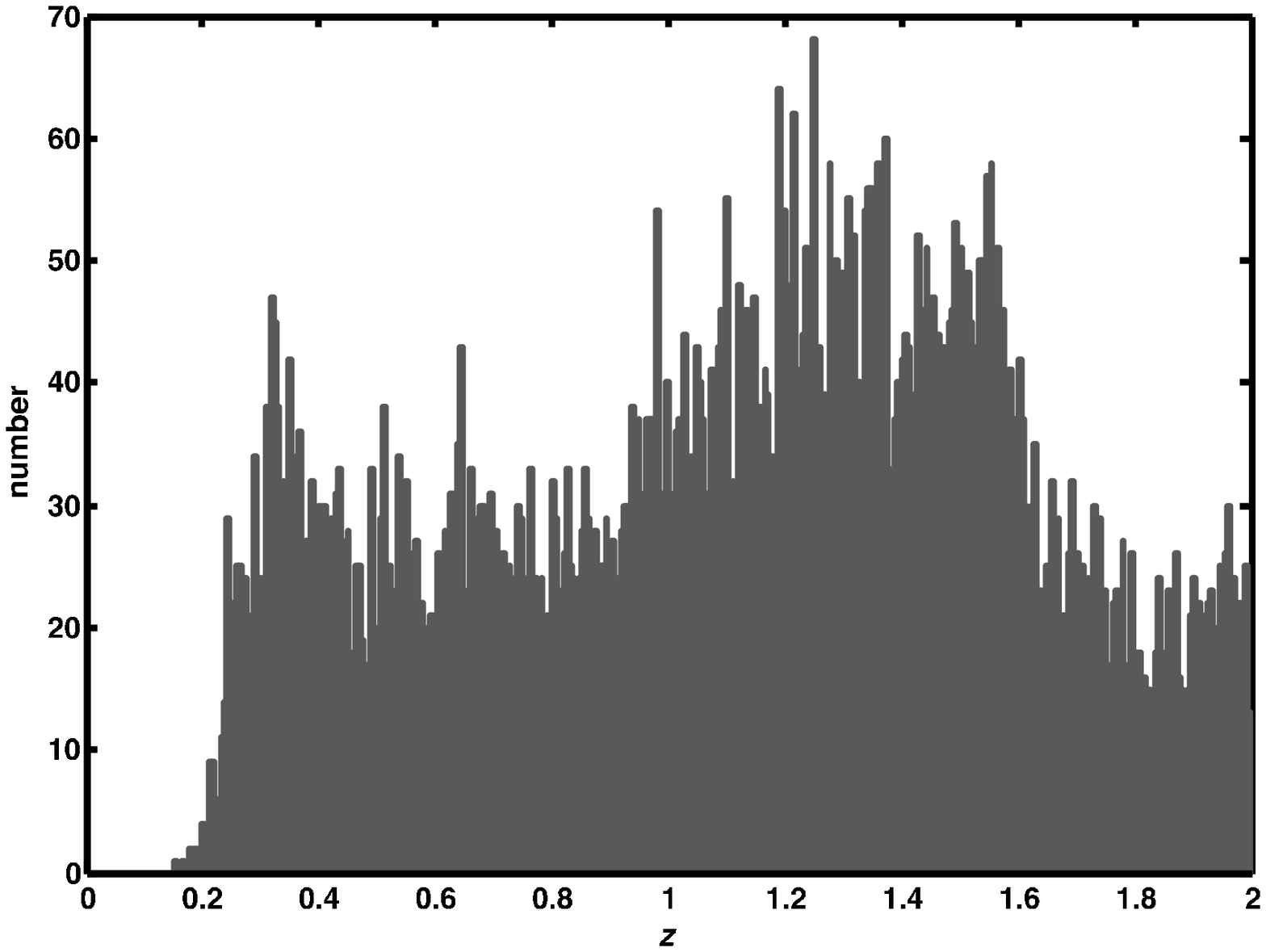}
 	\caption{\emph{Left:} Redshift distribution of N$_{sdss}$ quasars in bins of $\Delta z = 0.0046$. \emph{Right:} $i$-band magnitude distribution of N$_{sdss}$ quasars in bins of $	\Delta i = 0.0096$.  The y-axis represents the number of quasars in each bin. These quantities have been extracted from the SDSS DR5 catalog.}
     	\label{fig:sample_distros}
 	\end{centering}
     	\end{figure}

We further constrained this quasar catalog to the redshift range $z \le 2.0$. As described in \S \ref{data:mgii_sample}, the number density of intervening MgII systems is well-constrained only within this redshift range. This left us with a final sample size of N$_{sdss}$ = 11361 quasars. The $i$-band and redshift distributions of the SDSS sample can be found in Figure \ref{fig:sample_distros}.

\section{Analysis Methods} \label{methods} 
We produced a sample of simulated quasars that has the same statistical properties as the original SDSS sample. For every simulated quasar, we added a number of MgII absorbers whose characteristics are based on observationally-determined distributions to the sight-line. We accounted for the effect of dust by dimming the $i$-band magnitude of the quasar according to a SMC-type extinction law \citep{PREVOT,MENARD2008}. Evidence suggests that SMC-like dust is the most common form of dust associated with intervening MgII \citep{MENARD2008,KHARE2005,WILD2006}.

We applied 25 different input values for MgII $dn/dz$ in the range $0.24 \le (dn/dz)_{in} \le 0.90$. For a given $(dn/dz)_{in}$, we added quasars to the simulation until the number of quasars whose $i$-band magnitude was brighter than 19.1 equaled the number of quasars in the SDSS sample, N$_{sdss}$.  Once this stopping criterion was met, we were left with a sample of N$_{sim}$ simulated quasars.  Since extinction dimmed some of these quasars below 19.1, N$_{sim}$ was always larger than N$_{sdss}$. We interpret the quantity N$_{sim} - $N$_{sdss}$ as the number of quasars the SDSS misses due to dust dimming the $i$ magnitudes past the $i$=19.1 limit. We repeated this step 500 times and recorded the mean number of quasars in the simulation $\langle N_{sim}\rangle$, its associated uncertainty $\sigma_{N_{sim}}$, the mean output MgII number density $\langle (dn/dz)_{out}\rangle$, and its uncertainty $\sigma_{(dn/dz)_{out}}$, all for the given $(dn/dz)_{in}$. Finally, we quantified the relation between intrinsic and observed MgII number density and its associated uncertainty.

\section{Results} \label{results} 
\begin{figure}
    	\begin{centering}
     	\includegraphics[scale = .35]{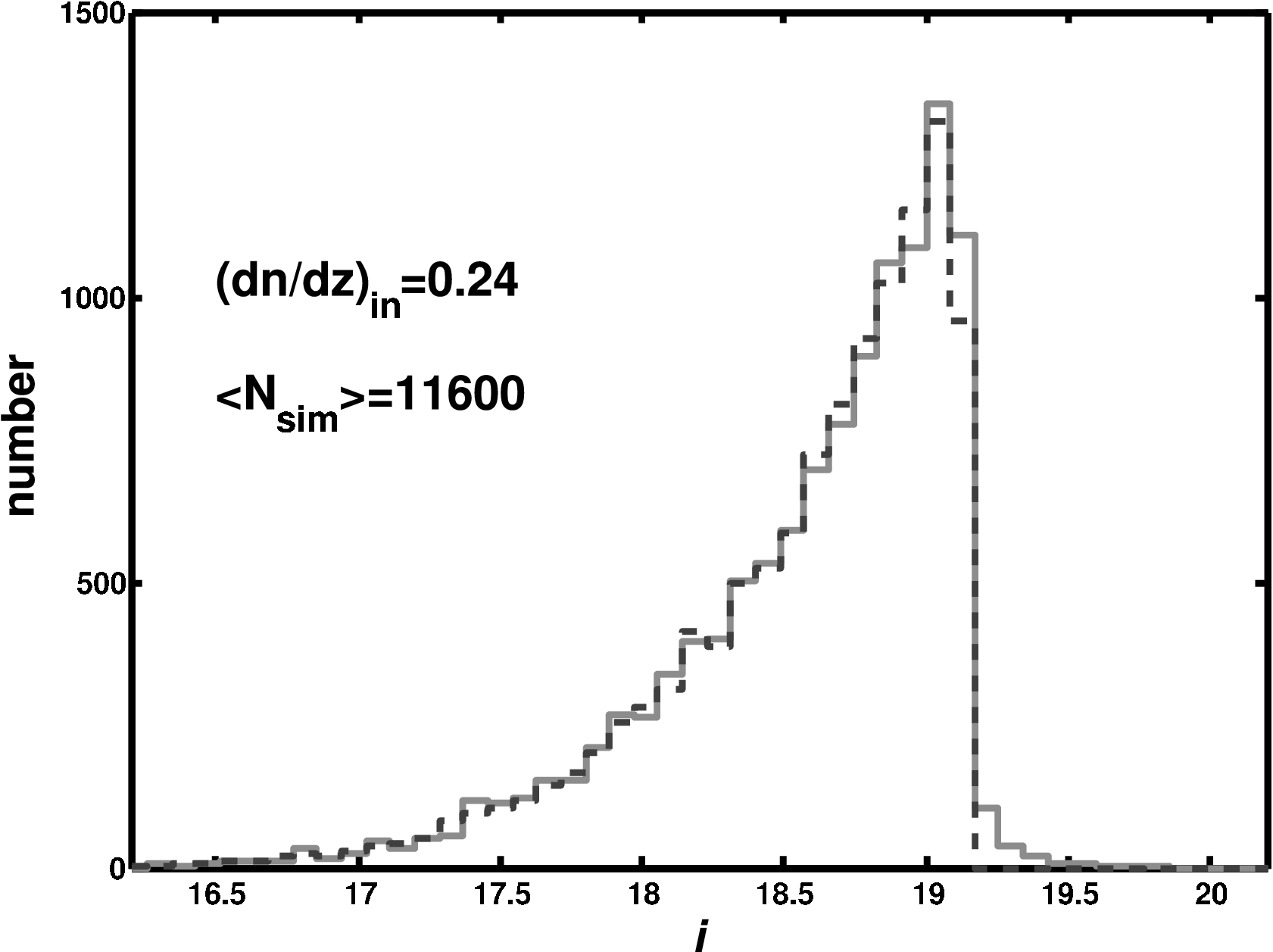}\includegraphics[scale=.35]{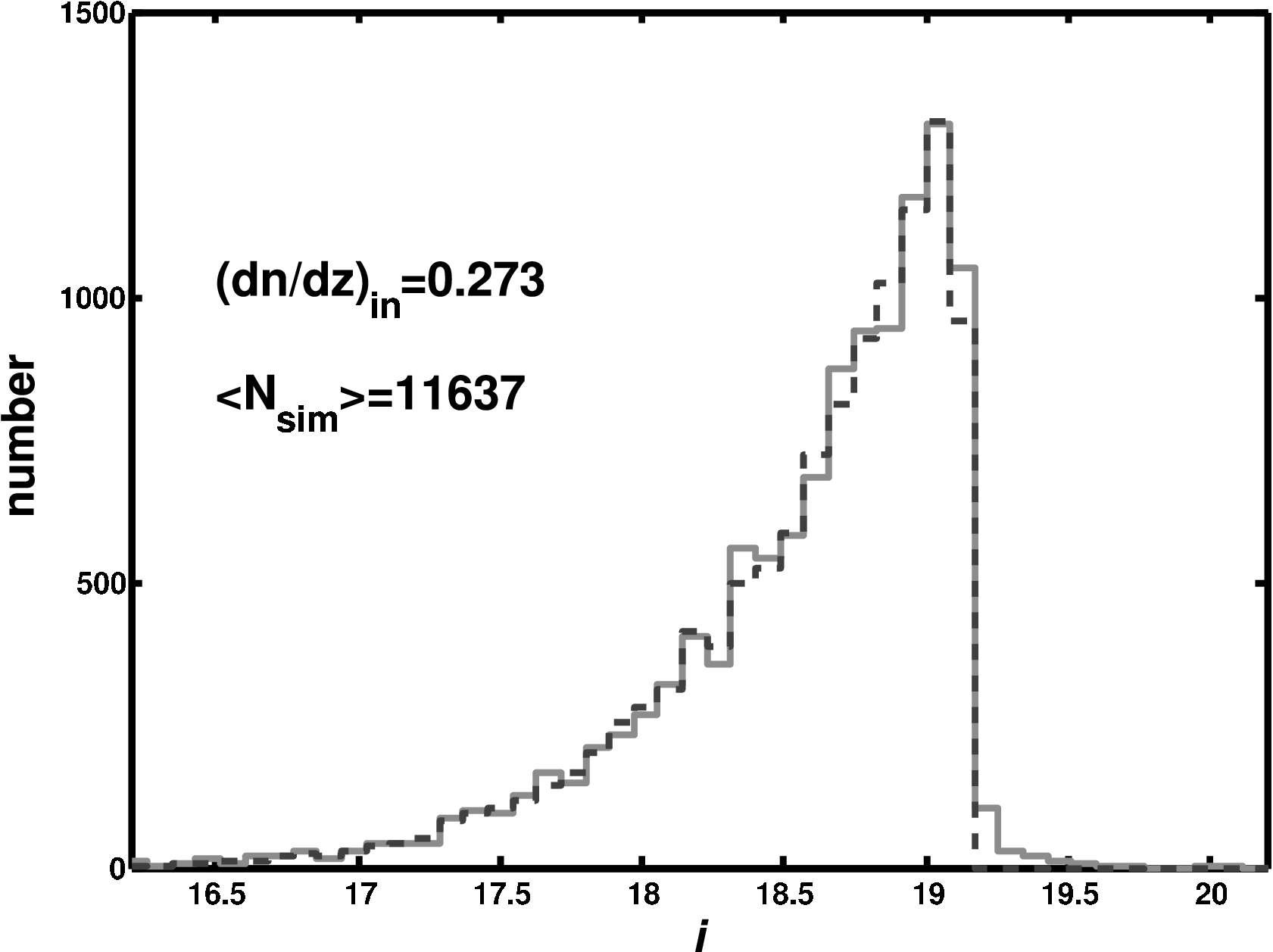}\includegraphics[scale=.35]{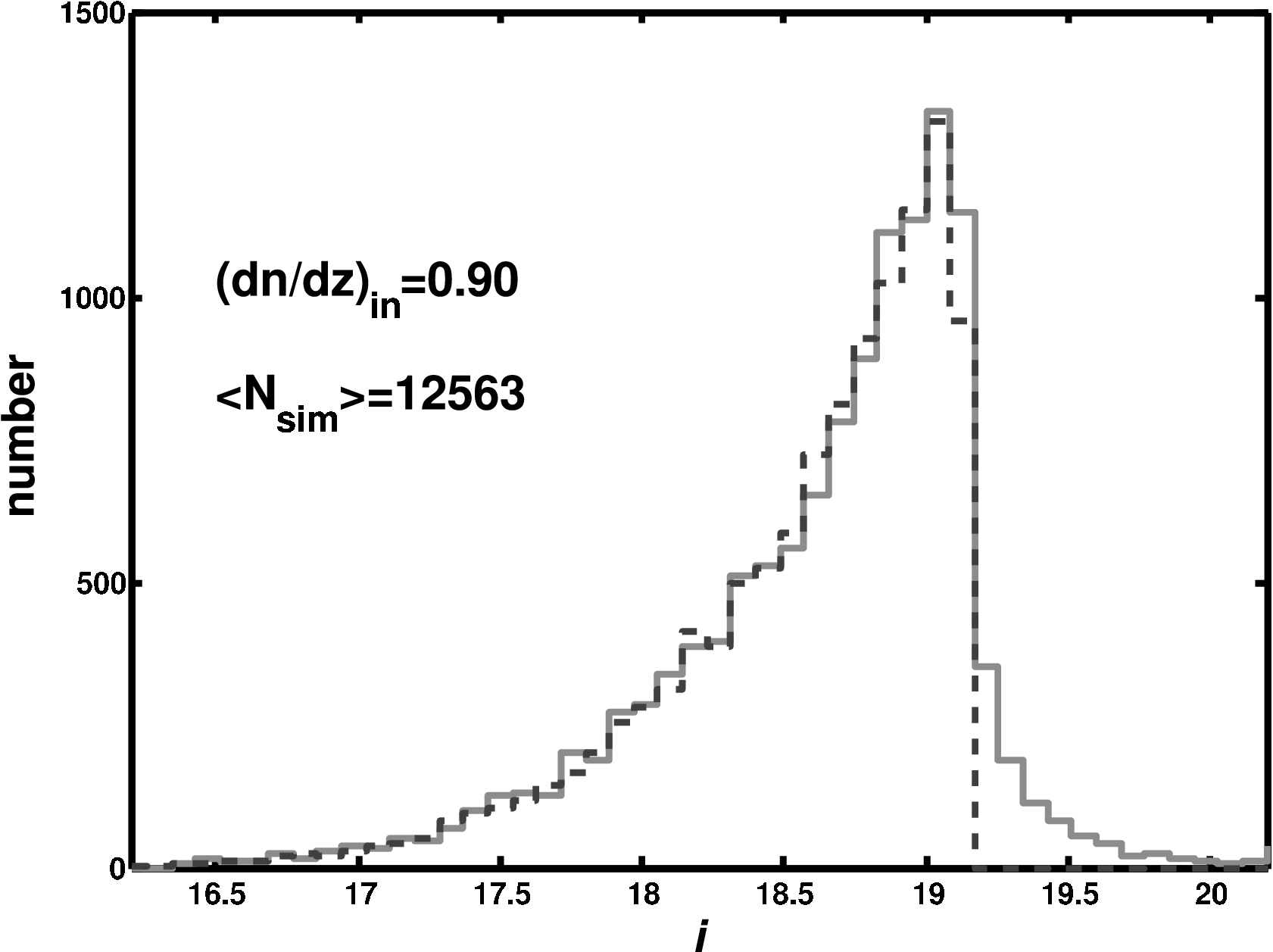}
 	\caption{\emph{Dashed line:} $i$-band magnitude distribution of the observed N$_{sdss}$ quasars. \emph{Solid line} $i$-band magnitude distribution of simulated N$_{sim}$ quasars. Example simulation outputs are plotted for $(dn/dz)_{in} = 0.24, 0.273,$ and 0.90, respectively.}
     	\label{fig:data-model_visualization}
 	\end{centering}
     	\end{figure}

For a given $(dn/dz)_{in}$, we created quasars until the stopping criterion described in \S \ref{methods} was met. The primary result from a single run of the simulation was a sample of quasar magnitudes. Several runs' samples for various values of $(dn/dz)_{in}$ are plotted in Figure \ref{fig:data-model_visualization} as a solid line histogram, along with the $i$-band distribution of quasars in the main SDSS catalog as a dashed line histogram. The same total number of quasars are brighter than 19.1 in each histogram. The solid bins to the right of $i$=19.1 represent the simulated quasars missed by the SDSS due to dust effects. 

\begin{figure} 
    	\begin{centering}
     	\includegraphics[scale = .75]{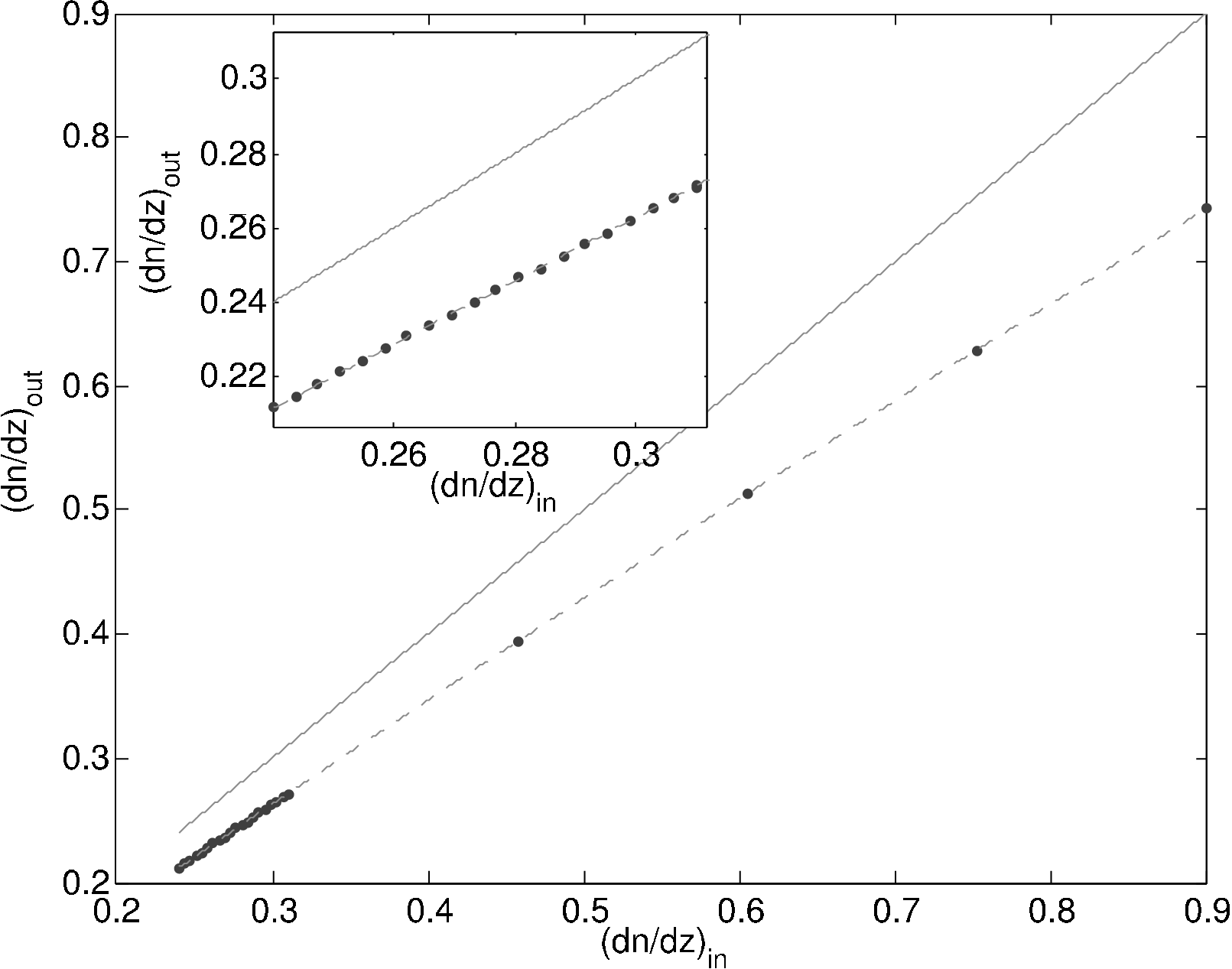}
 	\caption{$\langle (dn/dz)_{out} \rangle$ as a function of $(dn/dz)_{in}$, with the solid line representing equality. The data are well fit by a cubic polynomial. The inset highlights the data in the range $0.24 \lesssim (dn/dz)_{in} \lesssim 0.30$. $\langle (dn/dz)_{out} \rangle = 0.24$ when $(dn/dz)_{in} = 0.273$.}
     	\label{fig:ndensities}
 	\end{centering}
     	\end{figure}

From a single run with a given $(dn/dz)_{in}$, we derive the value for the MgII number density that a hypothetical observer would deduce from observing the populations of quasars that survive the $i<19.1$ cut ($(dn/dz)_{out}$). We show $\langle (dn/dz)_{out}\rangle$ as a function of $(dn/dz)_{in}$ in Figure \ref{fig:ndensities}, where the inset highlights the data in the $(dn/dz)_{in} \sim 0.25$ range. By interpolating the data with a cubic polynomial, we find that $\langle (dn/dz)_{out}\rangle = 0.24$ when $(dn/dz)_{in} = 0.273 \pm 0.002$. We estimate the uncertainty in this value by re-running the simulation with values of $(dn/dz)_{in}$ between 0.265 and 0.281 and finding the values for which $\langle (dn/dz)_{out}\rangle$ were different from 0.24 95\% of the time. 

\begin{figure}
    	\begin{centering}
     	\includegraphics[scale = .5]{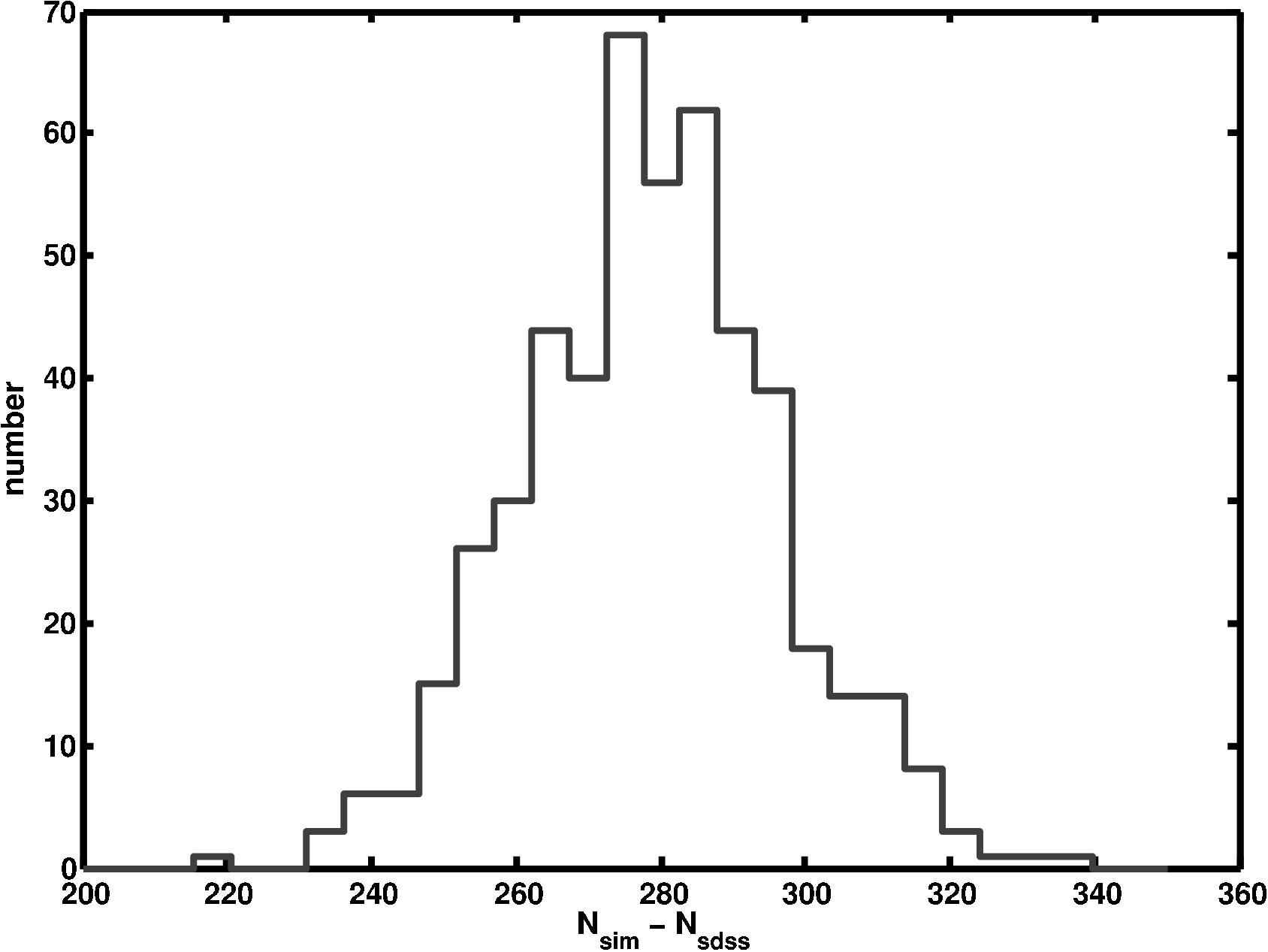}
 	\caption{Distribution of $N_{sdss}-N_{sim}$, which represent the number of quasars in the sample that would be observable if there were no dust extinction associated with MgII absorbers. The distribution is Gaussian, with $\langle N_{sim}-N_{sdss} \rangle = 280 \pm 20$ quasars.}
     	\label{fig:saved_n}
 	\end{centering}
     	\end{figure}

We present the distribution of $N_{sim} - N_{sdss}$ for an input number density of 0.273 in Figure \ref{fig:saved_n}. The data follow a Gaussian distribution, with $\langle N_{sim}-N_{sdss} \rangle = 280 \pm 20$ quasars. This implies that the SDSS misses $(2.4 \pm 0.2)\%$ of $z \le 2.0$ quasars due to dust extinction associated with intervening MgII absorbers.


\section{Discussion} \label{discussion} 
According to the simulation, the discrepancy between MgII number densities seen in GRB and quasar sight-lines cannot be fully explained by a selection bias introduced by dust extinction. We find that the observed incidence of intervening MgII absorbers in quasar sight-lines is lowered due to dust extinction by $\approx 10\%$: The unbiased MgII $dn/dz$ is likely to be 0.273, which is still very different from the GRB-observed value of $dn/dz=0.90$. Additionally, we find that at least $\approx 2\%$ of $z<2.0$ quasars are not included in the SDSS main quasar sample due to dust effects. This value is consistent with the estimate by \citet{MENARD2008} in which they use a different approach.

\citet{PORCIANI2007} have also tested explanations for the GRB and quasar MgII number density discrepancy. Most relevantly to this work, they use mock light-cones drawn from the Millennium Run \citep{SPRINGEL2005} to model the number of galaxies that would obscure a population of quasars at $z=2.3$, assigning MgII absorbers probabilistically to galaxies and assuming a relation between MgII and dust that best reproduces observed colors. Using this technique, they conclude that 16\% of quasars are missed by the SDSS due to dust effects, and that dust extinction can account for the GRB/quasar number density discrepancy only by a factor of $\sim 1.3 - 2$. A significant difference between this work and the work done by Porciani \etal\ is in how we calculate the dust traced by MgII. Porciani \etal\ use a Weibull distribution for color excess E(B-V) values, and assign these values to intervening MgII absorbers. A Weibull distribution is a continuous probability distribution whose shape is determined by two parameters. Besides constraining these two parameters with the quasar relative color excess $\Delta(g-i)$, the E(B-V) distribution is a free parameter in their model. In this work, we use an empirically derived relationship between MgII and dust reddening. Additionally, our method does not rely on the semi-analytical modeling in the Millennium Run.

Though a bias stemming from dust extinction is undoubtedly present in the SDSS, it only accounts for a small fraction of the observed MgII number density discrepancy between GRB and quasar sight-lines. Furthermore, none of the other proposed solutions to this problem are satisfactory: \citet{PONTZEN2007} show that there is no evidence for systematically smaller MgII equivalent widths over quasar broad emission regions (as compared to quasar continuum regions) and conclude that different MgII covering factors in the two types of lines of sight cannot explain the discrepancy. \citet{CUCCHIARA2008} investigate the properties of MgII absorbers in GRB and quasar lines of sight and find no difference, thereby arguing that the absorbers are correctly identified as intervening, non-intrinsic systems. \citet{PORCIANI2007} point out that a bias stemming from gravitational lensing requires that the quasar beam be larger than the GRB beam and that the optical depth of micro-lenses is likely significantly greater than observed. Both of these conditions, though possible, are unlikely given current observational data.

Although the MgII number density discrepancy is statistically significant, it is nevertheless important to increase the GRB absorber sample size to better constrain it. \citet{CHEN2009} have found galaxies within 2\arcsec of GRB hosts whose afterglows showed strong MgII absorbers. A comprehensive survey of these galaxies may also shed new light on this issue.

In summary, we have used a Monte-Carlo simulation to test if the factor of $\sim 4$ difference in MgII number densities observed in GRB and quasar sight-lines is due to dust extinction. We find that a dust extinction bias is likely to account for only $\approx 10\%$ of the observed overdensity in GRB sight-lines. Additionally, we estimate that $\approx 2\%$ of $z<2.0$ quasars are not included in the SDSS main sample because of dust obstruction. 

\acknowledgments

The authors acknowledge support from the Winslow Womack Research grant. The authors thank Brice M\'enard, Hsiao-Wen Chen, and the referee for useful comments. VS would like to acknowledge the support of the Guilford College Physics Department and the Max Planck Society.

Funding for the SDSS and SDSS-II has been provided by the Alfred P. Sloan Foundation, the Participating Institutions, the National Science Foundation, the U.S. Department of Energy, the National Aeronautics and Space Administration, the Japanese Monbukagakusho, the Max Planck Society, and the Higher Education Funding Council for England. The SDSS Web Site is http://www.sdss.org/.

The SDSS is managed by the Astrophysical Research Consortium for the Participating Institutions. The Participating Institutions are the American Museum of Natural History, Astrophysical Institute Potsdam, University of Basel, University of Cambridge, Case Western Reserve University, University of Chicago, Drexel University, Fermilab, the Institute for Advanced Study, the Japan Participation Group, Johns Hopkins University, the Joint Institute for Nuclear Astrophysics, the Kavli Institute for Particle Astrophysics and Cosmology, the Korean Scientist Group, the Chinese Academy of Sciences (LAMOST), Los Alamos National Laboratory, the Max-Planck-Institute for Astronomy (MPIA), the Max-Planck-Institute for Astrophysics (MPA), New Mexico State University, Ohio State University, University of Pittsburgh, University of Portsmouth, Princeton University, the United States Naval Observatory, and the University of Washington.

\bibliography{main_bibliography} 

\begin{thebibliography}{33}
\expandafter\ifx\csname natexlab\endcsname\relax\def\natexlab#1{#1}\fi

\bibitem[{{Adelman-McCarthy} {et~al.}(2008){Adelman-McCarthy}, {Ag{\"u}eros},
  {Allam}, {Allende Prieto}, {Anderson}, {Anderson}, {Annis}, {Bahcall},
  {Bailer-Jones}, {Baldry}, {Barentine}, {Bassett}, {Becker}, {Beers}, {Bell},
  {Berlind}, {Bernardi}, {Blanton}, {Bochanski}, {Boroski}, {Brinchmann},
  {Brinkmann}, {Brunner}, {Budav{\'a}ri}, {Carliles}, {Carr}, {Castander},
  {Cinabro}, {Cool}, {Covey}, {Csabai}, {Cunha}, {Davenport}, {Dilday}, {Doi},
  {Eisenstein}, {Evans}, {Fan}, {Finkbeiner}, {Friedman}, {Frieman},
  {Fukugita}, {G{\"a}nsicke}, {Gates}, {Gillespie}, {Glazebrook}, {Gray},
  {Grebel}, {Gunn}, {Gurbani}, {Hall}, {Harding}, {Harvanek}, {Hawley},
  {Hayes}, {Heckman}, {Hendry}, {Hindsley}, {Hirata}, {Hogan}, {Hogg}, {Hyde},
  {Ichikawa}, {Ivezi{\'c}}, {Jester}, {Johnson}, {Jorgensen}, {Juri{\'c}},
  {Kent}, {Kessler}, {Kleinman}, {Knapp}, {Kron}, {Krzesinski}, {Kuropatkin},
  {Lamb}, {Lampeitl}, {Lebedeva}, {Lee}, {Leger}, {L{\'e}pine}, {Lima}, {Lin},
  {Long}, {Loomis}, {Loveday}, {Lupton}, {Malanushenko}, {Malanushenko},
  {Mandelbaum}, {Margon}, {Marriner}, {Mart{\'{\i}}nez-Delgado}, {Matsubara},
  {McGehee}, {McKay}, {Meiksin}, {Morrison}, {Munn}, {Nakajima}, {Neilsen},
  {Newberg}, {Nichol}, {Nicinski}, {Nieto-Santisteban}, {Nitta}, {Okamura},
  {Owen}, {Oyaizu}, {Padmanabhan}, {Pan}, {Park}, {Peoples}, {Pier}, {Pope},
  {Purger}, {Raddick}, {Re Fiorentin}, {Richards}, {Richmond}, {Riess}, {Rix},
  {Rockosi}, {Sako}, {Schlegel}, {Schneider}, {Schreiber}, {Schwope}, {Seljak},
  {Sesar}, {Sheldon}, {Shimasaku}, {Sivarani}, {Smith}, {Snedden}, {Steinmetz},
  {Strauss}, {SubbaRao}, {Suto}, {Szalay}, {Szapudi}, {Szkody}, {Tegmark},
  {Thakar}, {Tremonti}, {Tucker}, {Uomoto}, {Vanden Berk}, {Vandenberg},
  {Vidrih}, {Vogeley}, {Voges}, {Vogt}, {Wadadekar}, {Weinberg}, {West},
  {White}, {Wilhite}, {Yanny}, {Yocum}, {York}, {Zehavi}, \&
  {Zucker}}]{SDSS_DR6_PAPER}
{Adelman-McCarthy}, J.~K. {et~al.} 2008, \apjs, 175, 297

\bibitem[{{Baker} {et~al.}(2000){Baker}, {Mathlin}, {Churches}, \&
  {Edmunds}}]{BAKER2000}
{Baker}, A.~C., {Mathlin}, G.~P., {Churches}, D.~K., \& {Edmunds}, M.~G. 2000,
  in Bulletin of the American Astronomical Society, Vol.~32, Bulletin of the
  American Astronomical Society, 1435--+

\bibitem[{{Cabanac} {et~al.}(2005){Cabanac}, {Hutsem{\'e}kers}, {Sluse}, \&
  {Lamy}}]{CABANAC2005}
{Cabanac}, R.~A., {Hutsem{\'e}kers}, D., {Sluse}, D., \& {Lamy}, H. 2005, in
  Astronomical Society of the Pacific Conference Series, Vol. 343, Astronomical
  Polarimetry: Current Status and Future Directions, ed. A.~{Adamson},
  C.~{Aspin}, C.~{Davis}, \& T.~{Fujiyoshi}, 498--+

\bibitem[{{Chen} {et~al.}(2009){Chen}, {Perley}, {Pollack}, {Prochaska},
  {Bloom}, {Dessauges-Zavadsky}, {Pettini}, {Lopez}, {Dall'aglio}, \&
  {Becker}}]{CHEN2009}
{Chen}, H.-W. {et~al.} 2009, \apj, 691, 152

\bibitem[{{Clowes} {et~al.}(1999){Clowes}, {Haines}, {Machura}, \&
  {Campusano}}]{CLOWES1999}
{Clowes}, R.~G., {Haines}, C.~P., {Machura}, I.~K., \& {Campusano}, L.~E. 1999,
  in Bulletin of the American Astronomical Society, Vol.~31, Bulletin of the
  American Astronomical Society, 1399--+

\bibitem[{{Cucchiara} {et~al.}(2008){Cucchiara}, {Jones}, {Charlton}, {Fox},
  {Einsig}, \& {Narayanan}}]{CUCCHIARA2008}
{Cucchiara}, A., {Jones}, T., {Charlton}, J.~C., {Fox}, D.~B., {Einsig}, D., \&
  {Narayanan}, A. 2008, ArXiv e-prints

\bibitem[{{Fan} {et~al.}(2006){Fan}, {Strauss}, {Becker}, {White}, {Gunn},
  {Knapp}, {Richards}, {Schneider}, {Brinkmann}, \& {Fukugita}}]{FAN2006}
{Fan}, X. {et~al.} 2006, \aj, 132, 117

\bibitem[{{Frank} {et~al.}(2007){Frank}, {Bentz}, {Stanek}, {Mathur},
  {Dietrich}, {Peterson}, \& {Atlee}}]{FRANK2007}
{Frank}, S., {Bentz}, M.~C., {Stanek}, K.~Z., {Mathur}, S., {Dietrich}, M.,
  {Peterson}, B.~M., \& {Atlee}, D.~W. 2007, \apss, 312, 325

\bibitem[{{Hamann} \& {Ferland}(1999)}]{HAMANN1999}
{Hamann}, F., \& {Ferland}, G. 1999, \araa, 37, 487

\bibitem[{{Hopkins} {et~al.}(2006){Hopkins}, {Somerville}, {Hernquist}, {Cox},
  {Robertson}, \& {Li}}]{HOPKINS2006}
{Hopkins}, P.~F., {Somerville}, R.~S., {Hernquist}, L., {Cox}, T.~J.,
  {Robertson}, B., \& {Li}, Y. 2006, \apj, 652, 864

\bibitem[{{Khare} {et~al.}(2005{\natexlab{a}}){Khare}, {Kulkarni}, {Lauroesch},
  {Fall}, {York}, {Welty}, {Crotts}, {Truran}, \& {Nakamura}}]{KHARE2005}
{Khare}, P. {et~al.} 2005{\natexlab{a}}, Bulletin of the Astronomical Society
  of India, 33, 219

\bibitem[{{Khare} {et~al.}(2005{\natexlab{b}}){Khare}, {York}, {vanden Berk},
  {Kulkarni}, {Crotts}, {Welty}, {Lauroesch}, {Richards}, {Alsayyad}, {Kumar},
  {Lundgren}, {Shanidze}, {Vanlandingham}, {Baugher}, {Hall}, {Jenkins},
  {Menard}, {Rao}, {Turnshek}, \& {Yip}}]{KHARE2005dust}
{Khare}, P. {et~al.} 2005{\natexlab{b}}, in IAU Colloq. 199: Probing Galaxies
  through Quasar Absorption Lines, ed. P.~{Williams}, C.-G. {Shu}, \&
  B.~{Menard}, 427--429

\bibitem[{{Lanzetta} \& {Bowen}(1990)}]{LANZETTA1990}
{Lanzetta}, K.~M., \& {Bowen}, D. 1990, \apj, 357, 321

\bibitem[{{Lanzetta} {et~al.}(1987){Lanzetta}, {Wolfe}, \&
  {Turnshek}}]{LANZETTA1987}
{Lanzetta}, K.~M., {Wolfe}, A.~M., \& {Turnshek}, D.~A. 1987, \apj, 322, 739

\bibitem[{{Mehlert} {et~al.}(2003){Mehlert}, {Noll}, \&
  {Appenzeller}}]{MEHLERT2003}
{Mehlert}, D., {Noll}, S., \& {Appenzeller}, I. 2003, \apss, 284, 437

\bibitem[{{M{\'e}nard} {et~al.}(2008){M{\'e}nard}, {Nestor}, {Turnshek},
  {Quider}, {Richards}, {Chelouche}, \& {Rao}}]{MENARD2008}
{M{\'e}nard}, B., {Nestor}, D., {Turnshek}, D., {Quider}, A., {Richards}, G.,
  {Chelouche}, D., \& {Rao}, S. 2008, \mnras, 385, 1053

\bibitem[{{M{\'e}nard} {et~al.}(2005){M{\'e}nard}, {Zibetti}, {Nestor}, \&
  {Turnshek}}]{MENARD2005}
{M{\'e}nard}, B., {Zibetti}, S., {Nestor}, D., \& {Turnshek}, D. 2005, in IAU
  Colloq. 199: Probing Galaxies through Quasar Absorption Lines, ed.
  P.~{Williams}, C.-G. {Shu}, \& B.~{Menard}, 86--91

\bibitem[{{Mesinger} \& {Haiman}(2007)}]{MESINGER2007}
{Mesinger}, A., \& {Haiman}, Z. 2007, \apj, 660, 923

\bibitem[{{Nestor} {et~al.}(2005){Nestor}, {Turnshek}, \& {Rao}}]{NESTOR2005}
{Nestor}, D.~B., {Turnshek}, D.~A., \& {Rao}, S.~M. 2005, \apj, 628, 637

\bibitem[{Peterson(1997)}]{PETERSON}
Peterson, B.~M. 1997, An Introduction to Active Galactic Nuclei (Cambridge
  University Press)

\bibitem[{{Pettini}(2001)}]{PETTINI2001}
{Pettini}, M. 2001, in ESA Special Publication, Vol. 460, The Promise of the
  Herschel Space Observatory, ed. G.~L. {Pilbratt}, J.~{Cernicharo}, A.~M.
  {Heras}, T.~{Prusti}, \& R.~{Harris}, 113--+

\bibitem[{{Pontzen} {et~al.}(2007){Pontzen}, {Hewett}, {Carswell}, \&
  {Wild}}]{PONTZEN2007}
{Pontzen}, A., {Hewett}, P., {Carswell}, R., \& {Wild}, V. 2007, \mnras, 381,
  L99

\bibitem[{{Porciani} {et~al.}(2007){Porciani}, {Viel}, \&
  {Lilly}}]{PORCIANI2007}
{Porciani}, C., {Viel}, M., \& {Lilly}, S.~J. 2007, \apj, 659, 218

\bibitem[{{Prevot} {et~al.}(1984){Prevot}, {Lequeux}, {Prevot}, {Maurice}, \&
  {Rocca-Volmerange}}]{PREVOT}
{Prevot}, M.~L., {Lequeux}, J., {Prevot}, L., {Maurice}, E., \&
  {Rocca-Volmerange}, B. 1984, \aap, 132, 389

\bibitem[{{Prochaska} {et~al.}(2007){Prochaska}, {Wolfe}, {Howk}, {Gawiser},
  {Burles}, \& {Cooke}}]{PROCHASKA2007}
{Prochaska}, J.~X., {Wolfe}, A.~M., {Howk}, J.~C., {Gawiser}, E., {Burles},
  S.~M., \& {Cooke}, J. 2007, \apjs, 171, 29

\bibitem[{{Prochter} {et~al.}(2006){Prochter}, {Prochaska}, {Chen}, {Bloom},
  {Dessauges-Zavadsky}, {Foley}, {Lopez}, {Pettini}, {Dupree}, \&
  {Guhathakurta}}]{PROCHTER}
{Prochter}, G.~E. {et~al.} 2006, \apjl, 648, L93

\bibitem[{{Springel} {et~al.}(2005){Springel}, {White}, {Jenkins}, {Frenk},
  {Yoshida}, {Gao}, {Navarro}, {Thacker}, {Croton}, {Helly}, {Peacock}, {Cole},
  {Thomas}, {Couchman}, {Evrard}, {Colberg}, \& {Pearce}}]{SPRINGEL2005}
{Springel}, V. {et~al.} 2005, \nat, 435, 629

\bibitem[{{Steidel} {et~al.}(1994){Steidel}, {Dickinson}, \&
  {Persson}}]{STEIDEL1994}
{Steidel}, C.~C., {Dickinson}, M., \& {Persson}, S.~E. 1994, \apjl, 437, L75

\bibitem[{{Sudilovsky} {et~al.}(2007){Sudilovsky}, {Savaglio}, {Vreeswijk},
  {Ledoux}, {Smette}, \& {Greiner}}]{SUDILOVSKY}
{Sudilovsky}, V., {Savaglio}, S., {Vreeswijk}, P., {Ledoux}, C., {Smette}, A.,
  \& {Greiner}, J. 2007, \apj, 669, 741

\bibitem[{{Tejos} {et~al.}(2007){Tejos}, {Lopez}, {Prochaska}, {Chen}, \&
  {Dessauges-Zavadsky}}]{TEJOS2007}
{Tejos}, N., {Lopez}, S., {Prochaska}, J.~X., {Chen}, H.-W., \&
  {Dessauges-Zavadsky}, M. 2007, \apj, 671, 622

\bibitem[{{Wild} {et~al.}(2006){Wild}, {Hewett}, \& {Pettini}}]{WILD2006}
{Wild}, V., {Hewett}, P.~C., \& {Pettini}, M. 2006, \mnras, 367, 211

\bibitem[{{York} {et~al.}(2000){York}, {Adelman}, {Anderson}, {Anderson},
  {Annis}, {Bahcall}, {Bakken}, {Barkhouser}, {Bastian}, {Berman}, {Boroski},
  {Bracker}, {Briegel}, {Briggs}, {Brinkmann}, {Brunner}, {Burles}, {Carey},
  {Carr}, {Castander}, {Chen}, {Colestock}, {Connolly}, {Crocker}, {Csabai},
  {Czarapata}, {Davis}, {Doi}, {Dombeck}, {Eisenstein}, {Ellman}, {Elms},
  {Evans}, {Fan}, {Federwitz}, {Fiscelli}, {Friedman}, {Frieman}, {Fukugita},
  {Gillespie}, {Gunn}, {Gurbani}, {de Haas}, {Haldeman}, {Harris}, {Hayes},
  {Heckman}, {Hennessy}, {Hindsley}, {Holm}, {Holmgren}, {Huang}, {Hull},
  {Husby}, {Ichikawa}, {Ichikawa}, {Ivezi{\'c}}, {Kent}, {Kim}, {Kinney},
  {Klaene}, {Kleinman}, {Kleinman}, {Knapp}, {Korienek}, {Kron}, {Kunszt},
  {Lamb}, {Lee}, {Leger}, {Limmongkol}, {Lindenmeyer}, {Long}, {Loomis},
  {Loveday}, {Lucinio}, {Lupton}, {MacKinnon}, {Mannery}, {Mantsch}, {Margon},
  {McGehee}, {McKay}, {Meiksin}, {Merelli}, {Monet}, {Munn}, {Narayanan},
  {Nash}, {Neilsen}, {Neswold}, {Newberg}, {Nichol}, {Nicinski}, {Nonino},
  {Okada}, {Okamura}, {Ostriker}, {Owen}, {Pauls}, {Peoples}, {Peterson},
  {Petravick}, {Pier}, {Pope}, {Pordes}, {Prosapio}, {Rechenmacher}, {Quinn},
  {Richards}, {Richmond}, {Rivetta}, {Rockosi}, {Ruthmansdorfer}, {Sandford},
  {Schlegel}, {Schneider}, {Sekiguchi}, {Sergey}, {Shimasaku}, {Siegmund},
  {Smee}, {Smith}, {Snedden}, {Stone}, {Stoughton}, {Strauss}, {Stubbs},
  {SubbaRao}, {Szalay}, {Szapudi}, {Szokoly}, {Thakar}, {Tremonti}, {Tucker},
  {Uomoto}, {Vanden Berk}, {Vogeley}, {Waddell}, {Wang}, {Watanabe},
  {Weinberg}, {Yanny}, \& {Yasuda}}]{SDSS_TECHPAPER}
{York}, D.~G. {et~al.} 2000, \aj, 120, 1579

\bibitem[{{York} {et~al.}(2006){York}, {Khare}, {Vanden Berk}, {Kulkarni},
  {Crotts}, {Lauroesch}, {Richards}, {Schneider}, {Welty}, {Alsayyad}, {Kumar},
  {Lundgren}, {Shanidze}, {Smith}, {Vanlandingham}, {Baugher}, {Hall},
  {Jenkins}, {Menard}, {Rao}, {Tumlinson}, {Turnshek}, {Yip}, \&
  {Brinkmann}}]{YORK2006}
---. 2006, \mnras, 367, 945

\end{thebibliography}
\bibliographystyle{apj}

\end{document}